\documentclass[%
 reprint,
superscriptaddress,
 amsmath,amssymb,
 aps,
prb,
]{revtex4-2}

\usepackage{graphicx}
\usepackage{dcolumn}
\usepackage{bm}
\usepackage{xcolor}

\begin{document}

\title{Layered Kagome Compound Na$_2$Ni$_3$S$_4$ with Topological Flat Band}

\author{Junyao Ye}
\affiliation{International Center for Quantum Materials, School of Physics, Peking University, Beijing 100871, China}
\author{Yihao Lin}
\affiliation{International Center for Quantum Materials, School of Physics, Peking University, Beijing 100871, China}
\author{Haozhe Wang}
\affiliation{Department of Chemistry, Michigan State University, MI, 48824, USA}
\author{Zhida Song}
\affiliation{International Center for Quantum Materials, School of Physics, Peking University, Beijing 100871, China}
\author{Ji Feng}
\affiliation{International Center for Quantum Materials, School of Physics, Peking University, Beijing 100871, China}
\author{Weiwei Xie}
\affiliation{Department of Chemistry, Michigan State University, MI, 48824, USA}
\author{Shuang Jia}
\email{gwljiashuang@pku.edu.cn}
\affiliation{International Center for Quantum Materials, School of Physics, Peking University, Beijing 100871, China}
\affiliation{Hefei National Laboratory, Hefei 230088, China}
\affiliation{Interdisciplinary Institute of Light-Element Quantum Materials and Research Center for Light-Element Advanced Materials, Peking University, Beijing 100871, China}
\date{\today}

\begin{abstract}

	We report structural and electronic properties of Na$_2$Ni$_3$S$_4$, a quasi-two-dimensional compound composed of alternating layers of [Ni$_3$S$_4$]$^{2-}$ and Na$^{+}$. The compound features a remarkable Ni-based kagome lattice with a square planar configuration of four surrounding S atoms for each Ni atom. 	Magnetization and electrical measurements reveal a weak paramagnetic insulator with a gap of about 0.5 eV. 
Our band structure calculation highlights a set of topological flat bands of the kagome lattice derived from the rotated d$_{xz}$-orbital with $C_\mathrm{3}$ + $T$ symmetry in the presence of crystal-field splitting.

\end{abstract}

\pacs{Valid PACS appear here}
\maketitle


\section{\label{sec:level1}INTRODUCTION}

Kagome lattice has garnered significant attention in the community of condensed matter physics since the concept was introduced by Syozi \cite{1951Statistics}. A kagome lattice is a two-dimensional lattice pattern consisting of corner-sharing triangles, which is characterized by a large degree of geometrical frustration, and as a result, it can bear exotic quantum states such as quantum spin liquid~\cite{freedman2010site,10.1038/nature11659}. On the other hand, a kagome lattice typically generates the electronic structure characterized by a Dirac point (DP) at the Brillouin zone (BZ) corner, a saddle point at the BZ boundary, and a flat band (FB) over the whole BZ, which are the source for nontrivial band topology~\cite{Ghimire2020}. With the inclusion of magnetism and spin-orbit coupling (SOC), the electrons can realize a strongly interacting topological phase in a kagome lattice~\cite{PhysRevLett.128.096601}. Recent studies have revealed various quantum physical properties in transition-metal-based kagome materials. For example, charge density wave (CDW) order and superconductivity were observed in a family of vanadium-based layered kagome materials AV$_3$Sb$_5$ (A = K, Rb, Cs)~\cite{2021Unconventional,yang2020giant,PhysRevB.104.L041103}. Spectrum meaturements revealed the existence of the massive Dirac Fermion on the transition-metal-based kagome lattices in Fe$_3$Sn$_2$~\cite{PhysRevLett.121.096401} and TbMn$_6$Sn$_6$~\cite{2020Quantum} which generate significant berry curvature field and anomalous Hall effect.

The geometric frustration of a kagome lattice can cause destructive quantum interference, leading to a dispersionless FB in the momentum space. Such FB is prone to various instabilities aroused by electron-electron interactions, which can lead to various exotic many-body phenomena, such as superconductivity~\cite{cao2018a,zhaohe2021}, fractional quantum anomalous Hall~\cite{xuxiaodong2023},  ferromagnetism~\cite{jiang2019,PhysRevLett.82.4312,tasaki1992}, excitonic insulator~\cite{sethi2021} and Wigner crystals~\cite{wu2007,chen2018}. The FBs can have flatness-protected touching points with dispersive band, leading to various topological phases~\cite{liuzheng2014}, including Chern insulators~\cite{xu2015,rhim2019}, fractional quantum Hall insulators~\cite{sheng2011,tang2011,neupert2011,sun2011}, topological insulators~\cite{Yoshinori2013,guo2009b,chang2013}  and Weyl semimetals~\cite{zhou2019} when they are subjected to time-reversal symmetry (TRS) breaking and SOC.

Although correlated topological phases have been observed in Moire mini-FBs in twisted bilayer graphene and transition metal dichalcogenides~\cite{cao2018a,cao2018b,LIU2021100085}, the FB still needs to be explored in bulk kagome materials. The dispersionless FB in kagome lattice requests a perfect phase cancellation, in which  an isotropic nearest-neighbor single-particle hopping interaction is critical. Besides the simplest kagome lattice model assuming an s-orbital per lattice site, the physical underpinning for the existence of ideal FB in the transition-metals based kagome lattice is largely unknown. Most materials found in the laboratory involve multiple orbitals that defy such simplistic theoretical models. As a result, the observed FBs in many transition-metal-based kagome metals feature significant dispersion due to the character of the contributing orbitals and the accompanying anisotropic inter-orbital hopping~\cite{kang2020,liu2020, liu2023,PhysRevX.13.041049}. 

To search for the potential FB in kagome compounds, we synthesize a new layered Ni-based kagome compound Na$_2$Ni$_3$S$_4$. This compound belongs to a group of compounds having the chemical formula of A$_2$T$_3$Ch$_4$, where A is the alkali metal including Li, Na, K, and Rb, T is a divalent transition metal such as Ni, Pd, Zn, and Mn, and Ch is S and Se~\cite{10.1002/zaac.19915970105, 10.1002/chin.197502007, HUSTER1974254, 10.1016/0925-8388(95)02064-0, 10.1002/zaac.19966220409, 10.1016/S0022-3697(03)00139-2, PhysRevB.69.045103, 10.1016/S0925-8388(01)01738-8, 10.1016/S0304-8853(97)00515-5, 10.1143/JPSJ.68.3668, hasegawa2004raman}. These compounds obey the 18-electron rule and are highly insulative. They crystallize in various layered structures which are built up by stacking [T$_3$Ch$_4$]$^{2-}$ and A$^+$ layers alternatively~\cite{10.1016/0925-8388(95)02064-0}. Most of A$_2$T$_3$Ch$_4$ compounds crystallize in the Cs$_2$Zn$_3$S$_4$-type structure in which the [T$_3$Ch$_4$]$^{2-}$ layer consists of edge-sharing TCh$_4$ tetrahedra~\cite{hasegawa2004raman}. On the other hand, when the transition metal is Ni and Pd, the compounds crystallize in orthorhombic structures in which the [Ni$_3$Ch$_4$]$^{2-}$ and [Pd$_3$Ch$_4$]$^{2-}$ layers consist of distorted double layers Ch-based honeycomb lattice and Ni(Pd)-based kagome lattice~\cite{10.1143/JPSJ.68.3668}. Each Ni(Pd) atom is located at the center of four Ch atoms in a rectangle planer configuration. The single crystals of these quasi-two-dimensional compounds can be easily mechanically cleaved.

The A$_2$Ni(Pd)$_3$Ch$_4$ compounds are all highly insulating with a band gap of about 1 eV~\cite{Hondou2005, 10.1016/S0925-8388(01)01738-8, 10.1016/S0022-3697(03)00139-2, 10.1016/S0304-8853(97)00515-5, 10.1143/JPSJ.68.3668, 10.1016/0925-8388(95)02064-0}.
Because the divalent nickel and palladium cations manifest a low spin configuration $3d^8$ with $S$ being 0, they in general demonstrate weak diamagnetic or paramagnetic signal~\cite{10.1143/JPSJ.68.3668}. However, it was reported that  the Rb$_2$Ni$_3$S$_4$ specimens have a small residual ferromagnetic moment if they were soak in water~\cite{Hondou2005}. This unexpected ferromagnetism was attractive because it may be related to a long-sought flat-band ferromagnetism characteristic of a kagome lattice~\cite{10.1143/PTP.99.489}.
A recent study observed metal-insulator transition and superconductivity in Rb$_2$Pd$_3$Se$_4$ under high pressure~\cite{li2022superconductivity}.
The electronic structures of  A$_2$Ni(Pd)$_3$Ch$_4$ compounds are highlighted by a manifold of narrow bands just below the Fermi level $E_\mathrm{F}$ which are derived from the hybridization between the Ni(Pd) $d$ and Ch $p$ states ~\cite{PhysRevB.69.045103}.
This manifold of bands has very narrow bandwidth and sharp density of states and therefore they were regarded as dispersionless FBs in the literature~\cite{HONDOU20071815}.
However the orbital characterization and energy dispersion of these bands remain largely unknown.
The characteristic topological FBs of the kagome lattice have not been observed in the electronic states of A$_2$Ni(Pd)$_3$Ch$_4$.

In this research, we report the structural and electronic properties of a new compound Na$_2$Ni$_3$S$_4$ which built up by staking [Ni$_3$S$_4$]$^{2-}$ and multiple Na$^{+}$ layers alternatively.
Unlike the other A$_2$Ni(Pd)$_3$Ch$_4$ compounds, the Ni-based kagome lattice has no structural distortion and therefore a perfect phase cancellation is plausible.
Physical property measurements reveal it as a week paramagnetic insulator with a band gap about 0.5 eV.
Moreover, the electronic band structure calculation shows there is a set of highly two-dimensional, three-band manifold derived from $d_{xz}$-orbital below the Fermi level.
This manifold contains an FB with a touching point with the dispersive band along the $\varGamma-A$ direction, a van Hove singularity, and a Dirac crossing point, all characterized for a kagome lattice.
The perfect FB is formed due to the rotation of d$_{xz}$-orbital on the kagome lattice caused by the crystal field splitting which conforms to the $C_\mathrm{3}$ + $T$ symmetry \cite{liu2023}.
Such coordination of Ni atoms satisfies the exacting destructive interference which is rarely observed in real materials.

\section{\label{sec:level2} METHODS}

The process of synthesizing crystals of Na$_2$Ni$_3$S$_4$ involves a carbonate-flux method \cite{10.1143/JPSJ.68.3668} in which Na$_2$CO$_3$ (99$\%$), Ni (99.98$\%$), and S (99.999$\%$) were mixed in a molar ratio of $5:1:12$ and loaded into an alumina crucible.
The mixture was then heated in a tube furnace with flowing argon gas in which the temperature was gradually increased to 890 $^\circ$C at a rate of 10 $^\circ$C/min. After 180 minutes, the furnace was turned off and yellow chunks containing crystals with mm in size yielded. The chunks can be dissociated into water and the flake-shaped crystals were obtained. 

\begin{figure}[h]
	\begin{center}
		\includegraphics[clip, width=0.5\textwidth]{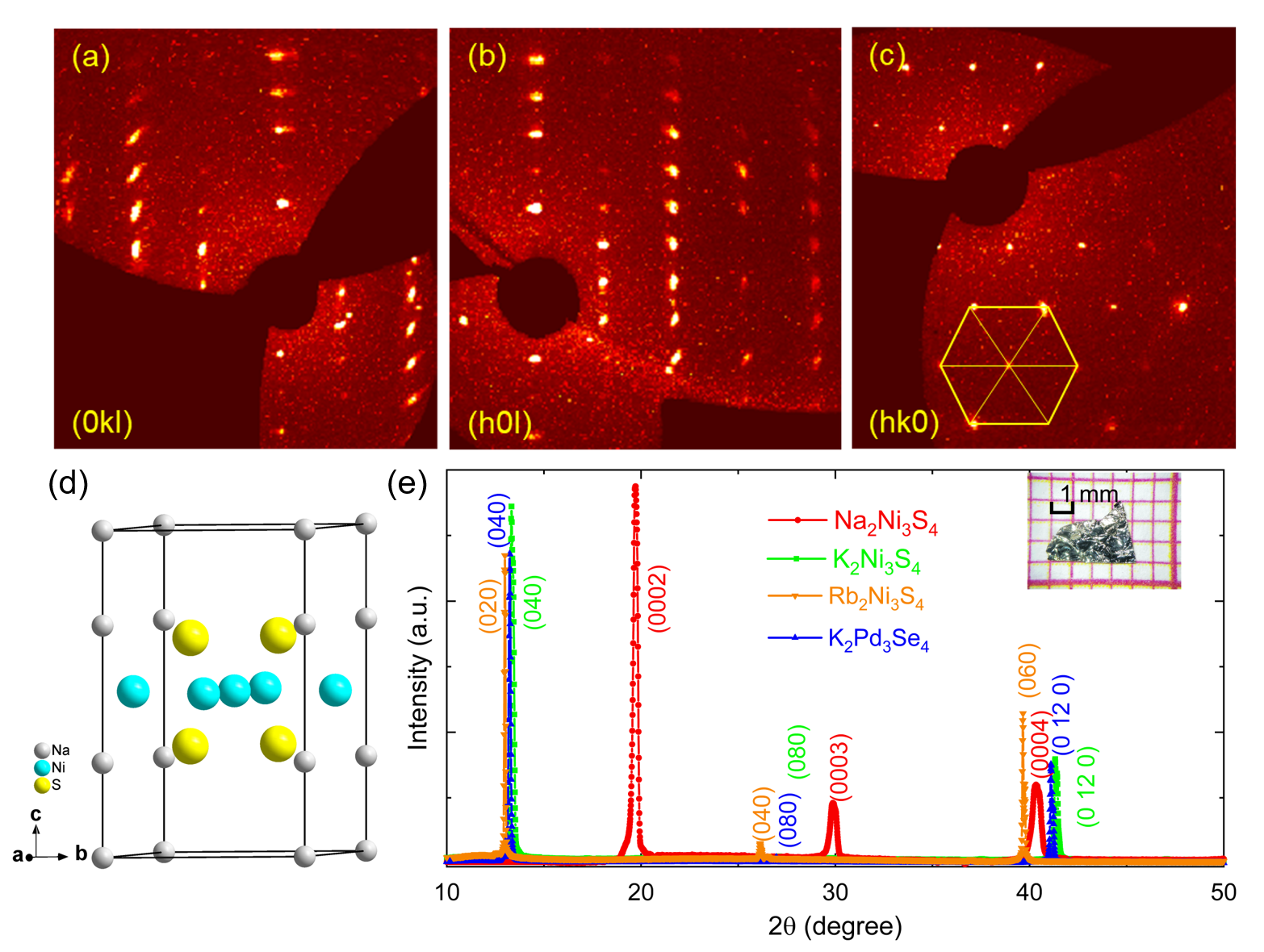}\\[1pt] 
		\caption{Precession images calculated from single-crystal X-ray diffraction data (a) (0kl) plane, (b) (h0l) plane, and (c) (hk0) plane. The resolved spots are consistent with space group $P6/mmm$. (d) The schematic crystal structure of Na$_2$Ni$_3$S$_4$ (one unit cell). (e) The X-ray diffraction pattern of flakes of A$_2$T$_3$Ch$_4$ (The inset is a photo of Na$_2$Ni$_3$S$_4$ flake). }
		\label{fig:1}
	\end{center}
\end{figure}

Single-crystal X-ray diffraction (XRD) measurement using Bruker D8 Eco-Quest Diffractometer with Mo radiation ($\lambda_{K_\alpha}$ = 0.71073 \AA) was carried out at room temperature. Several crystals with the size about ~50 $\mu$m$\times$ 50 $\mu$m $\times$ 20 $\mu$m) were picked up and tested.
Figure \ref{fig:1} (a), (b), and (c) depict precession images derived from the diffraction data in different directions.
The frames were integrated with the Bruker SAINT software package using a narrow-frame algorithm. The integration of the data using a hexagonal unit cell in the space group of $P6/mmm$, which are based upon the refinement of the XYZ-centroids. Data were corrected for absorption effects using the Multi-Scan method (SADABS). The crystal structure of Na$_2$Ni$_3$S$_4$, as shown in Fig. \ref{fig:1}(d), was refined using the direct method and full-matrix least-squares on the F$^2$ model with the SHELXTL software. The cell constants of Na$_2$Ni$_3$S$_4$ are $a = b = 5.656(5)$ \AA; $c = 8.930(12)$ \AA, $V = 247.4(6)$ \AA$^3$. Table I provides comprehensive details on the crystal structure of Na$_2$Ni$_3$S$_4$.

A Quantum Design magnetic property measurement system (MPMS-3) was used to perform the magnetization measurement for several samples with the applied field direction being along the $c$-axis at different temperatures.  Electric transport measurements for the single crystals were carried out in Quantum Design physical property measurement system (PPMS-9). Standard four-point method was adopted and gold paste was used as the contacts.

Density-functional theory (DFT) calculations were performed for the experimental determined hexagonal unit cell, as shown in Table I, using Vienna ab-initio Simulation Package (VASP v6.3.0)\cite{kresse1996d,kresse1996e}. The calculations are carried out within generalized gradient approximation(GGA) using PBE-type functional and recommended projector augmented-wave (PAW) pseudopotential\cite{blochl1994b}. Plane wave energy cutoff is chosen to be 270 eV and an $8\times8\times8$ $\Gamma$-centered $\mathbf{k}$-mesh is used. Orbital characters of Kohn-Sham orbitals are calculated under the PAW framework.

\section{\label{sec:level3}RESULTS AND ANALYSIS}

To demonstrate the unique structure of Na$_2$Ni$_3$S$_4$, we have performed X-ray diffraction measurements on several flakes of single-crystalline A$_2$T$_3$Ch$_4$ whose largest nature surface was attached in the pucks in Rigaku Mini-flux 600 using a Cu-K$_\alpha$ radiation source.
As shown in Fig. \ref{fig:1}(e), the positions of the diffraction peaks represent the stacking spaces in the quasi-two-dimensional A$_2$T$_3$Ch$_4$ compounds.
The space $d$ is about $8.93 \AA$ for Na$_2$Ni$_3$S$_4$, which aligns with the $c$ value obtained in the single-crystal X-ray diffraction.
On the other hand, the $d$ values are $6.80 \AA$, $6.54 \AA$ and $6.50 \AA$ for Rb$_2$Ni$_3$S$_4$, K$_2$Pd$_3$Se$_4$ and K$_2$Ni$_3$S$_4$, respectively.
This result reveals that the spacing of the Ni$_3$S$_4$ layer is the largest in the A$_2$T$_3$Ch$_4$ compounds, indicating a distinct stacking way which will be discussed below.

\begin{center}
	{\footnotesize{\bf Table 1.} Single crystal structure refinement data for Na$_2$Ni$_3$S$_4$.\\
		\vspace{2mm}
		\begin{tabular}{cc}
			\hline
			{\bf Empirical formula} & Na$_2$Ni$_3$S$_4$ \\\hline
			Formula weight (g/mol) & 350.32  \\
			Temperature & 300(2)  \\
			Crystal system & Hexagonal  \\
			Space group; Z & $P6/mmm$; 1  \\
			$a = b$ (\AA) & 5.656(5)  \\
			$c$ (\AA)  & 8.930(12)  \\
			Volume (\AA$^3$) & 247.4(6) \\
			Extinction coefficient & 0.7(4)  \\
			Theta range ($^{\circ}$) & 2.281 to 33.149  \\
			No. independent reflections & 222  \\
			No. parameters & 15  \\
			$R_{1} : {\omega}R_{2}(I>2{\delta}(I))$ & 0.2309 : 0.4881  \\
			R indices (all data) $R_{1} : {\omega}R_{2}$ & 0.2503 : 0.5147  \\
			Goodness-of-fit on F$^2$ & 1.992  \\
			\hline
	\end{tabular}}
\end{center}

\begin{center}
	{\footnotesize{\bf Table 2.} Atomic coordinates and equivalent isotropic displacement parameters of Na$_2$Ni$_3$S$_4$  system (U$_eq$ is defined as one-third of the trace of the orthogonalized U$_{ij}$ tensor ({\AA}$^2$)).\\
		\vspace{2mm}
		\begin{tabular}{ccccccc}
			\hline
			\bf Atom & \bf Wyck & \bf x & \bf y & \bf z & \bf Occ & \bf U${_eq}$  \\
			\hline
			Ni & 3$g$ & 1/2 & 0 & 1/2 & 1 & 0.021(3)  \\
			S & 4$h$ & 1/3 & 1/3 & 0.3343(2) & 1 & 0.028(3)  \\
			Na1 & 2$e$ & 0 & 0 & 0.290 & 0.667 & 0.032(9)  \\
			Na2 & 1$a$ & 0 & 0 & 0 & 0.667 & 0.028(9)  \\
			\hline
	\end{tabular}}
\end{center}

\begin{figure}[htbp]
	\begin{center}
		\includegraphics[clip, width=0.5\textwidth]{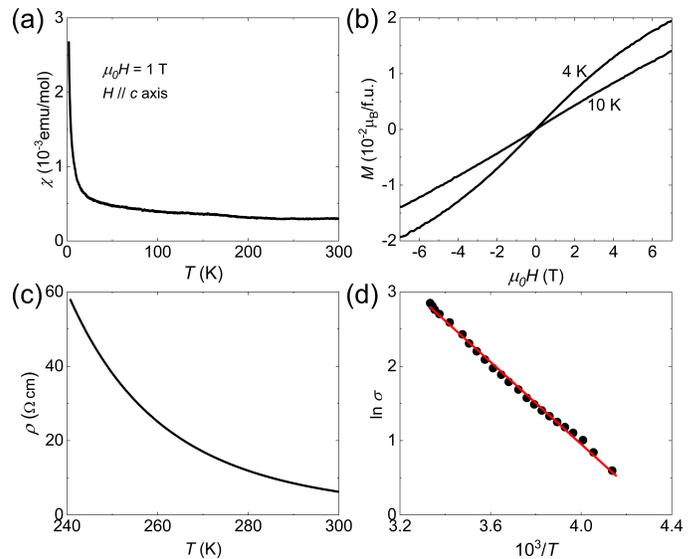}\\[1pt] 
		\caption{ (a) Temperature dependence of magnetic susceptibility of Na$_2$Ni$_3$S$_4$. An external magnetic field $\mu{_0}H$ = 1 T was applied along the crystallographic $c$-axis. (b) Field dependence of magnetization at 4 K and 10 K. (c) Temperature dependence resistivity of Na$_2$Ni$_3$S$_4$. (d) Fitting of the logarithm of conductance versus the reciprocal of temperature.
		}
		\label{fig:2}
	\end{center}
\end{figure}

Figure \ref{fig:2}(a) shows the temperature dependence of magnetic susceptibility for several pieces of Na$_2$Ni$_3$S$_4$ in a temperature range from 300 to 2 K when an external magnetic field $\mu{_0}H$ = 1 T was applied along the crystallographic $c$-axis. Na$_2$Ni$_3$S$_4$ exhibits temperature-independent paramagnetic signal at high temperature region and a Curie tail at low temperature region. 
We fit the susceptibility data below 30~K by using a constant term $\chi_0$ plus a Curie term $\frac{C}{T}$.
The fitting data are $\chi_0$ = 3.6 $\times$ 10$^{-4}$ emu/mol and $C$ = 0.004 emu K/mol, leading to an effective moment being 0.14 $\mu_B$ per formula unit, which likely comes from some magnetic impurity.
Figure \ref{fig:2}(b) shows the field dependence of magnetization up to 7 T at 4 K and 10 K. One can see a strict linear relation between $M$ and $H$ at 10~K while the $M$ curve at 4~K seems to be the linear background plus a paramagnetic contribution which is saturated at about 4~T. 
These observations confirm that the magnetization of the sample consists of a small Curie term of the localized moments and a temperature-independent paramagnetic term, which are likely due to extrinsic sources, such as lattice imperfections and impurity.
We notice the magnetization of Na$_2$Ni$_3$S$_4$ is distinct from the weak ferromagnetization of Rb$_2$Ni$_3$S$_4$ whose origin is not clear at this point~\cite{10.1016/S0925-8388(01)01738-8}.

Figure \ref{fig:2}(c) shows a semiconductor-like, temperature dependence of the resistivity in Na$_2$Ni$_3$S$_4$ from 300 K to 240 K.
The values of the resistivity are comparable to those for Rb$_2$Ni$_3$S$_4$~\cite{10.1016/S0925-8388(01)01738-8} and below 240 K they are beyond the capacity of our instrument.
As shown in Fig. \ref{fig:2}(d), an energy of activation  $E_g$  is estimated as $0.48$ eV  by using the formula of $\sigma = A e^{(-E_g/2k_\mathrm{B}T)}$, which is close to the value of Rb$_2$Ni$_3$S$_4$ at high temperatures~\cite{10.1016/S0925-8388(01)01738-8}.

\begin{figure*}[htbp]
	\begin{center}
		\includegraphics[clip, width=1\textwidth]{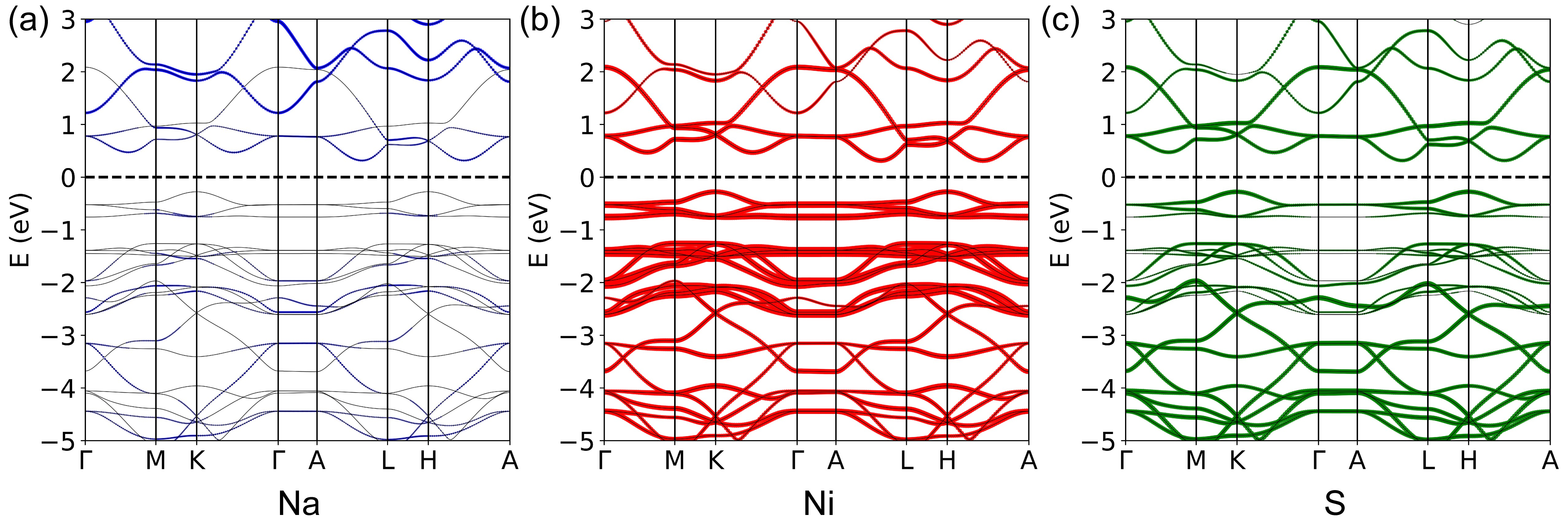}\\[1pt] 
		\caption{Atomic projected band structure of Na$_2$Ni$_3$S$_4$ for atoms (a) Na, (b) Ni, and (c) S.}
		\label{fig:3}
	\end{center}
\end{figure*}

In order to better understand the electronic structure of Na$_2$Ni$_3$S$_4$, we utilized a DFT calculation in generalized gradient approximation. To simplify the calculation, we assume that Na atoms are located at the $2e$ position, which preserves the average structural symmetry $P6/mmm$.
Figure ~\ref{fig:3} (a), (b) and (c) are the atomic projected band structure with respect to Na, Ni and S respectively.
Na hardly contributes to the energy band within the range from 0 to -5 eV. On the other hand, Ni and S have strong hybridization in the calculated energy range. 
The electronic structure has an indirect band gap of 0.75 eV with no state around $E_F$ , which is slightly larger than the estimation on the temperature dependent resistivity.
Below $E_F$ there is a narrow three-band manifold which is isolated with the deeper bands with a gap about 0.6 eV.
This three-band manifold is somehow similar as what observed in the band structures of Rb$_2$Ni$_3$S$_4$ \cite{PhysRevB.69.045103} and Rb$_2$Pd$_3$Se$_4$~\cite{li2022superconductivity} but the bandwith for Na$_2$Ni$_3$S$_4$ (0.5 eV) is more narrow.

\begin{figure}[htbp]
	\begin{center}
		\includegraphics[clip, width=0.5\textwidth]{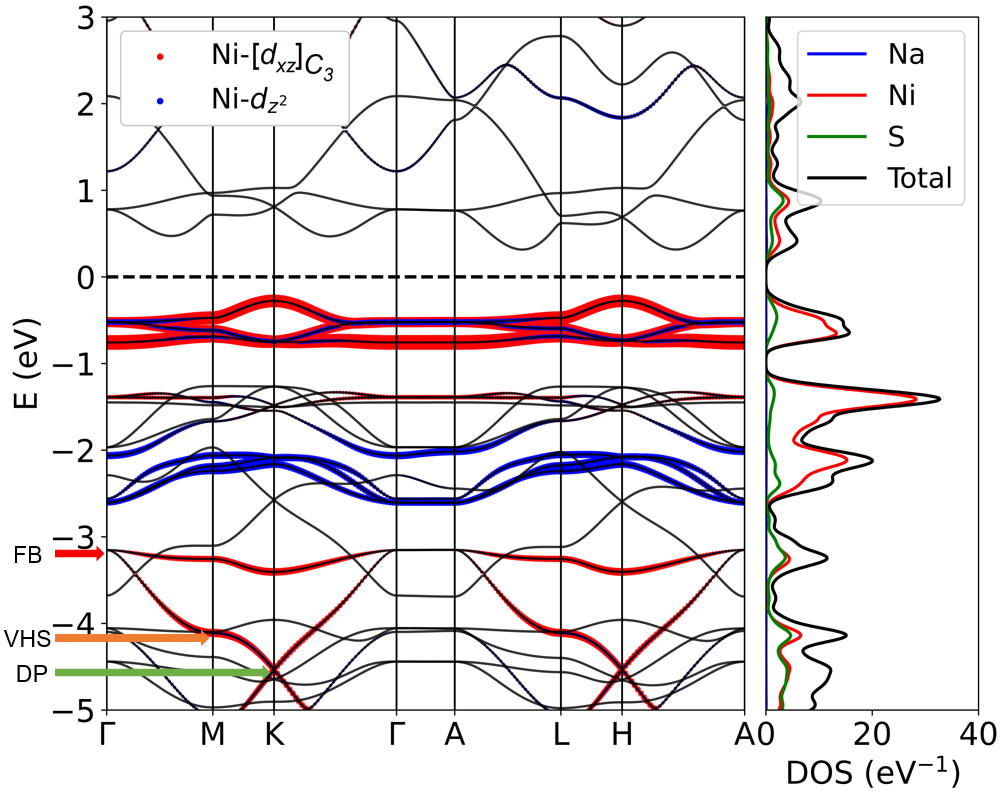}\\[1pt] 
		\caption{ Band structure (left) and atomic projected density of state (right) of Na$_2$Ni$_3$S$_4$ calculated by DFT. Band resolved projected density of state of $C_\mathrm{3}$ + $T$ symmetry-adapted d$_{xz}$ orbitals is highlighted in band structure as red, and d$_{z^2}$ orbital as blue.}
		\label{fig:4}
	\end{center}
\end{figure}

\begin{figure}[htbp]
	\begin{center}
		\includegraphics[clip, width=0.5\textwidth]{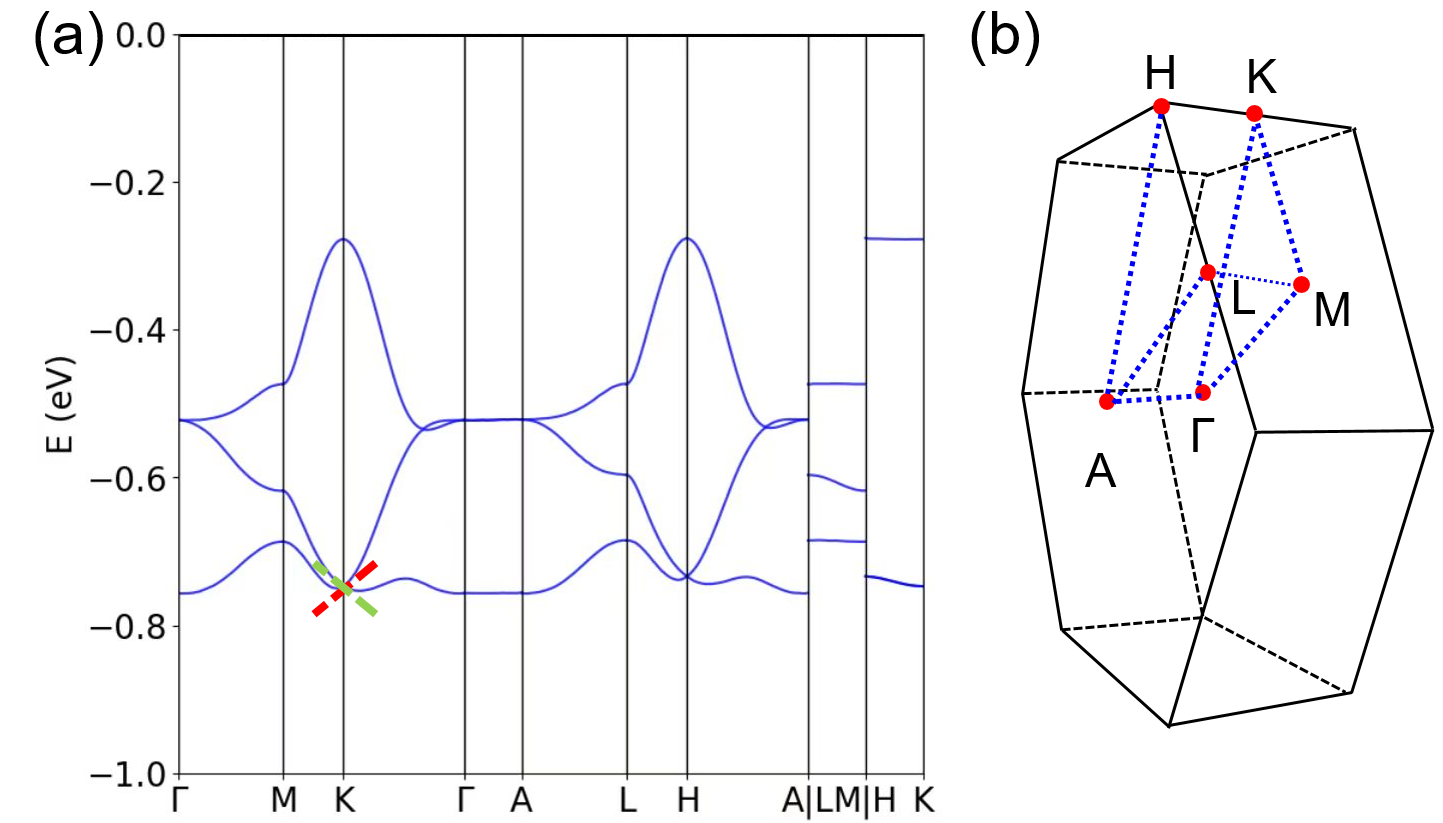}\\[1pt] 
		\caption{Calculated electronic energy dispersion of Na$_2$Ni$_3$S$_4$ in the range of 0 to -1.0 eV.}
		\label{fig:5}
	\end{center}
\end{figure}

Because the Ni atoms are located in a square planar coordination of sulfur with the site symmetry of $D_{2h}$, the five-fold degenerate $d$-orbitals of Ni will split into 5 manifolds under the orthorhombic crystal electric field.
We take into account the local $xyz$ coordination for the square plane of Ni atom and analyze the atomic $d$ orbitals as shown in Fig. \ref{fig:4}.
The three-band manifold has $d_{xy}$ and $d_{z^2}$ character.
To clarify whether there exist topological FB in the three-band manifold, we zoomed in the electronic energy dispersion in the range of 0 to -1 eV in Fig. \ref{fig:5}.
The apparently FB at -0.75 eV indeed comes from two bands which cross at K and H points.
The touching point in K is a linear-dispersed Dirac point ubiquitous in the kagome lattice.
This is clear that the characteristic topological FB of kagome lattice is absent in this three-band manifold just below $E_F$.

On the other hand, there is another highly two-dimensional, three-band manifold derived form $d_{xz}$-orbital between -3 and -5 eV without other $d$-orbital.
This manifold contains an FB at about -3.2 eV with the bandwith less than 0.2 eV, a van Hove singularity at -4.0 eV, and a Dirac crossing point at the K and H points at -4.5 eV.
The FB touches with the dispersive band along the $\varGamma-A$ direction.
All these features are characterized for a kagome lattice\cite{liuzheng2014} which is absent in the electronic structure for the other A$_2$T$_3$Ch$_4$ compounds.

\section{\label{sec:level3}DISCUSSION}

\begin{figure}[h]
	\begin{center}
		\includegraphics[clip, width=0.5\textwidth]{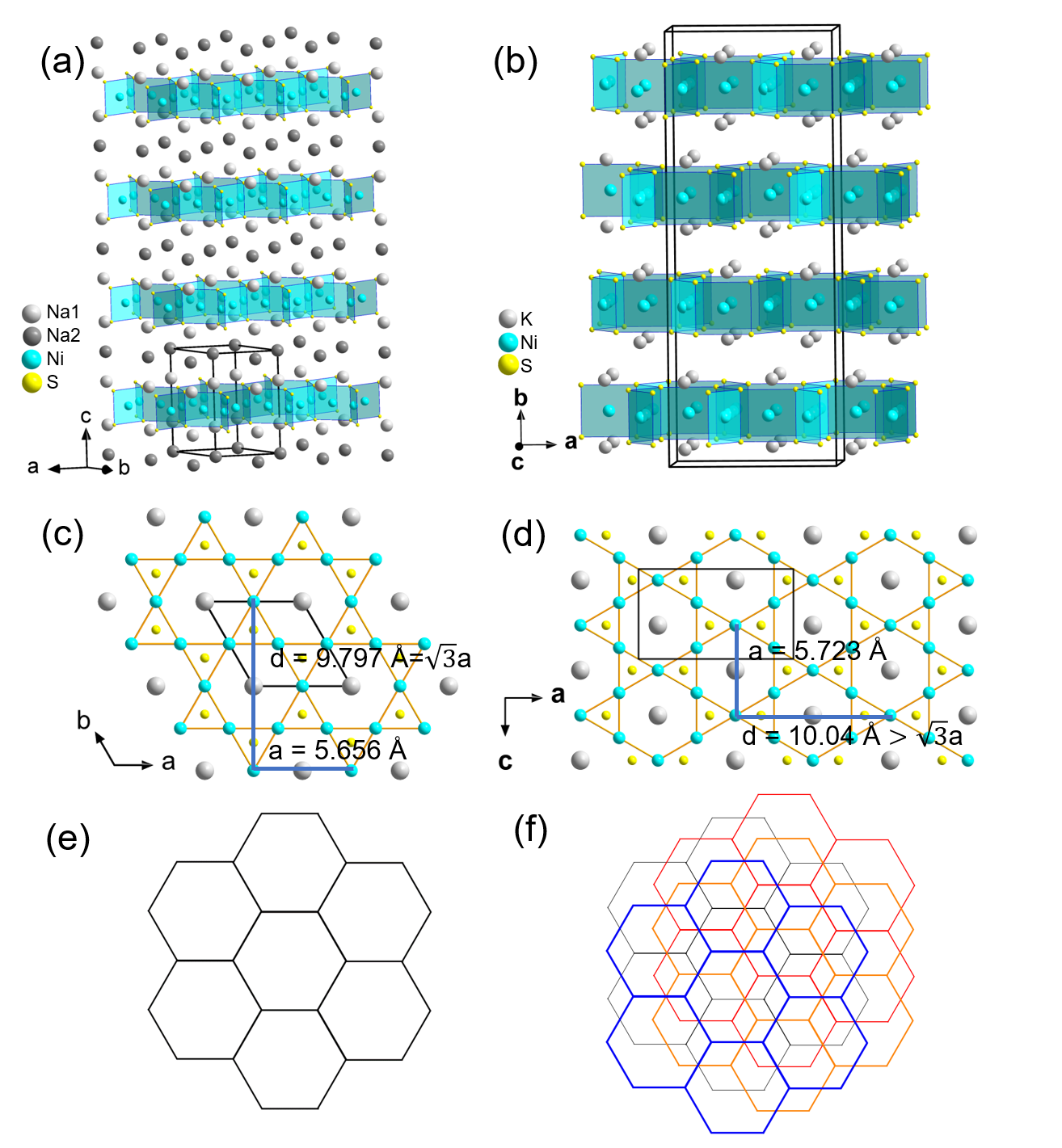}\\[1pt] 
		\caption{(a), (b) Schematic diagram of the crystal structure of the Na$_2$Ni$_3$S$_4$ and K$_2$Ni$_3$S$_4$. (c) A Ni-based kagome layer in Na$_2$Ni$_3$S$_4$ ($P6/mmm$), and (d) A distorted Ni-based kagome layer in K$_2$Ni$_3$S$_4$ ($Fddd$). (e) The stacking of the S-based honeycomb lattices in Na$_2$Ni$_3$S$_4$. (f) The stacking of the S-based honeycomb lattices in K$_2$Ni$_3$S$_4$.}
		\label{fig:6}
	\end{center}
\end{figure}

We now comparably demonstrate the crystal structures of Na$_2$Ni$_3$S$_4$ and K$_2$Ni$_3$S$_4$ as representative for the other A$_2$Ni(Pd)$_3$Ch$_4$ compounds in Fig. \ref{fig:6}.
The most significant difference is that the Na atoms occupy 2/3 of the two different crystallographic sites equivalently in Na$_2$Ni$_3$S$_4$ whereas the alkli atoms fully occupy only one crystallographic site in the other A$_2$Ni(Pd)$_3$Ch$_4$ compounds.
In Na$_2$Ni$_3$S$_4$, the Na1 site is located upper and down the center of the six sulfur rings from the Ni$_3$S$_4$ slab while the Na2 site is sandwiched between the two Na1 site along the $c$ axis.
The existence of this additional Na$^{+}$ layer makes a significant large spacing ($8.93 \AA$), 40\% larger than that in other A$_2$T$_3$Ch$_4$ compounds with no such site.
On the other hand, the staggered stacking of the Ni$_3$S$_4$-K-K-Ni$_3$S$_4$ layers leads to a quadruple $b$ axis in K$_2$Ni$_3$S$_4$~\cite{10.1016/0925-8388(95)02064-0}, making the sulfur honeycomb in the Ni$_3$S$_4$ layer repeat itself in every four layers, as shown in Fig. \ref{fig:6} (f).

The structure of the Ni$_3$S$_4$ layer also has pronounced difference between Na$_2$Ni$_3$S$_4$ and the other A$_2$Ni(Pd)$_3$Ch$_4$ compounds.
In Na$_2$Ni$_3$S$_4$, the Ni$_3$S$_4$ layer is formed by the linked edge-sharing NiS$_4$ square plane in such a way as to define two honeycomb network of S and one kagome network of Ni in between.
This NiS$_4$ square plane reads the Ni-S distance being about $2.205 \AA$ and the S-Ni-S angles being 84.478$^\circ$ and 95.522$^\circ$, close to a perfect sulfur square planar coordination.
The Ni-based kagome lattice has no distortion and the Ni-Ni distance is $2.828(3) \AA$ equivalently, which is about 6{\%} higher than the Ni-Ni contacts in nickel metal~\cite{10.1016/0925-8388(95)02064-0}, indicating strong atomic interactions between the Ni atoms. 

In the other A$_2$Ni(Pd)$_3$Ch$_4$ compounds, the Ni$_3$S$_4$ layer is stretched along the $c$ direction, leading to two crystallographic sites for the Ni atoms.
As the result, the Ni-S bondlengthes have about 0.4\% difference for the two Ni sites in the NiS$_4$ square plane.
The S-Ni-S bond angles for the two Ni sites range from 82.572$^\circ$ to 97.447$^\circ$ and 82.908$^\circ$ to 98.399$^\circ$, respectively~\cite{10.1016/0925-8388(95)02064-0}.
The distorted Ni-based kagome is characterized by different Ni-Ni distance, ranging from 2.862 \AA~to 2.889 \AA.

The perfect Ni-based kagome lattice seems to exist in Na$_2$Ni$_3$S$_4$ uniquely for all A$_2$T$_3$Ch$_4$ compounds.
As pointed out by Ref. ~\cite{10.1002/zaac.19915970105},  large alkali metal ions can cause lattice distortion in these layer structural compounds.
Nevertheless, the existence of the additional Na+ layer plays a crucial role for forming the perfect kagome lattice which is critical for the existence of a toplogical FB.

\begin{figure}[htbp]
	\begin{center}
		\includegraphics[clip, width=0.5\textwidth]{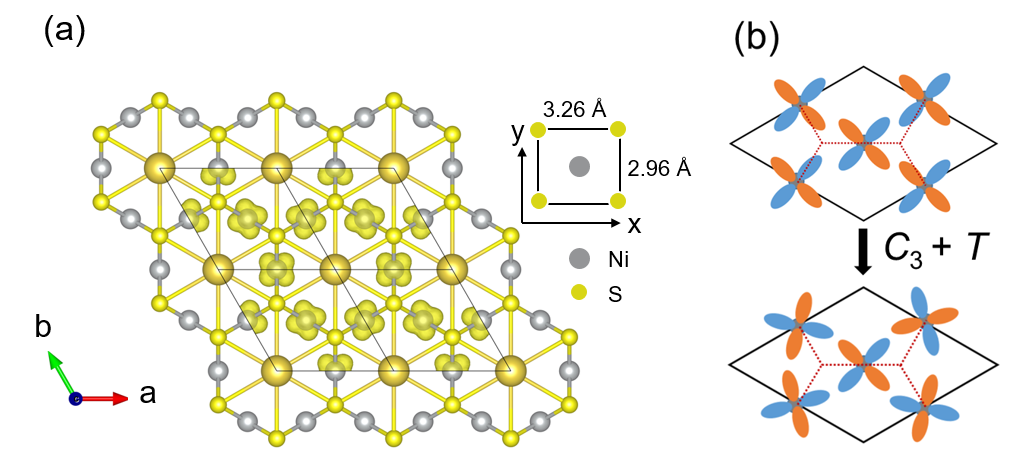}\\[1pt] 
		\caption{ (a) The charge density of the FBs immediately below the Fermi level manifests the character of $C_3$ + $T$ symmetry-adapted d$_{xz}$ orbitals. The upper right inset is a schematic diagram of the rectangular coordination environment of Ni-S in Na$_2$Ni${_3}$S$_4$. (b) Illustration of d$_{xz}$ orbital with $C_\mathrm{3}$ + $T$ rotation to produce a FB. The black and red curves represent bands without and with rotation, respectively.}
		\label{fig:7}
	\end{center}
\end{figure}

In a kagome lattice, the perfect destructive interference often relies on a tight-binding model of $s$-orbitals, requiring a single rotational invariant orbital on each site with isotropic hopping~\cite{liu2023}.
Obliviously a distorted kagome lattice with two sets of Ni-Ni distances, such as those in the other A$_2$T$_3$Ch$_4$ compounds, cannot own isotropic hopping and perfect destructive interference.
For the d$_{xy}$, d$_{x^2-y^2}$, d$_{xz}$, and d$_{yz}$ orbitals,  Kim \it et.~al. \rm pointed out that the properly rotated $d$-orbital can form perfect FBs on kagome lattice if they conform to the underlying kagome lattice symmetry, most notably $C_\mathrm{3}$ rotation plus translation ($C_\mathrm{3}$ + $T$)~\cite{liu2023}. The resultant FB will have the same shape as that of the $s$-orbital model, possessing a quadratic touching point with another dispersive band.


The above theoretical analysis help us to understand why the topological FB of $d_{xz}$ orbitals remain intact in Na$_2$Ni$_3$S$_4$.
The crystal electric field of the square planar coordination of sulfur orients in such direction of the kagome lattice that the properly rotated $d_{xz}$ orbital preserve the three-fold rotation symmetry $C_\mathrm{3}$ + $T$ of the crystals.
As shown in Fig. \ref{fig:7}(a), the partial charge at each site has the shape of $d_{xz}$ and manifests the $d_{xz}$ orbital character on each Ni atom in its orthorhombic local coordinate. 
Therefore the highly two-dimensional, three-band manifold purely derived form $d_{xz}$-orbital between -3 and -5 eV generate a set of ideal topological FB of kagome lattice.
On the other hand, the d$_{z^2}$ orbitals need a weak $\pi$ bonding to hopping inter-site.
The hybridization between the $d_{xz}$ and d$_{z^2}$ orbitals interrupts perfect destructive interference and therefore the topological FB is absent in the three manifold just below $E_F$.

\section{\label{sec:level4}CONCLUSION}

In summary, we have synthesized a new layered kagome compound Na$_2$Ni$_3$S$_4$ which demonstrates novel crystal structure, highlighted by a large layer-layer spacing and perfect Ni-based kagome lattice.
The calculated band structure shows that there is a set of three-band manifold derived from $d_{xz}$-orbital between -3 and -5 eV which contains FB, Dirac point and saddle piont, all characterized by the kagome lattice.
Because the formation of FBs in Na$_2$Ni$_3$S$_4$ is due to the rotation of d$_{xz}$-orbital which conforms to the $C_\mathrm{3}$ + $T$ symmetry, such construction promises a perfect destructive interference which is rarely observed in real materials.

\section*{ACKNOELEDGMENTS}
We would like to thank Wenlong Ma, Xiaoxiao Zhang and Jiahao Yang for helpful discussions. This work was supported by the National Natural Science Foundation of China (Grants No.12141002, No. 12225401), the National Key Research and Development Program of China (Grant No. 2021YFA1401902), the CAS Interdisciplinary Innovation Team, and the Strategic Priority Research Program of Chinese Academy of Sciences (Grant No. XDB28000000).

\bibliography{Na2Ni3S4}

%

\clearpage

\end{document}